# Fluctuation-theory constraint for extensive entropy in Monte-Carlo simulations

Ralph V. Chamberlin, Department of Physics, Arizona State University,
Tempe AZ 85287-1504
George H. Wolf, Department of Chemistry and Biochemistry, Arizona State University,
Tempe AZ 85287-1604

Abstract

The entropy per particle in most Monte-Carlo simulations is size dependent due to correlated energy fluctuations. Guided by nanothermodynamics, we find a constraint for the Ising model that enhances the fluctuations and lowers the free energy, while making the entropy homogeneous, additive, and extensive. Although the average interaction energy becomes size dependent, the resulting distribution of energies provides a mechanism for the heterogeneity found in the dynamics of many materials.

Corresponding Author: Ralph V. Chamberlin, Department of Physics, Box 871504, Arizona State University, Tempe, AZ 85287-1504 USA

Phone: +1-480-965-3922 Fax: +1-480-965-7954 ralph.chamberlin@asu.edu

The size dependence of thermal variables provides an important test of self consistency in thermodynamics. In macrothermodynamics there are two distinct categories of variables: intensive variables (e.g. temperature and magnetic field) that are independent of system size; and extensive variables (e.g. entropy and magnetization) that are linearly proportional to the number of particles in the system. For small systems, one or more of the extensive variables may become nonextensive, where the variable no longer depends linearly on system size. Here we use Monte-Carlo simulations to study how local interactions cause nonextensive behavior in the standard 3-D Ising model. We avoid simple surface terms by studying subsystem "blocks" of $n$ particles, inside a large system with periodic boundary conditions and size $N>>n$, so that the main nonextensivity comes from thermal fluctuations. Although it is well known that most Monte-Carlo simulations of such blocks in the Ising model have energy fluctuations per particle $<(\Delta U)^2>/n$ that decrease with decreasing $n$ [1], now we focus on the fact that when integrated between two temperatures the nonextensive heat capacity yields a nonextensive entropy. Specifically, the heat capacity of two separate blocks is less than the heat capacity of the same two blocks combined into a single block. Although superficially similar to Gibbs' paradox for combining systems of indistinguishable particles [2], here the nonadditive entropy comes from interparticle interactions [3,4].

Two diverse scenarios have been used to modify thermodynamics when the interaction length becomes comparable to system size. Tsallis statistics starts with a generalized definition of entropy that is intrinsically nonextensive [5,6], which may accommodate the behavior found in standard Monte-Carlo simulations; but some experts remain skeptical about the physical foundation of this alternate entropy [7]. Hill's small-system thermodynamics [8,9] starts with the postulate that entropy is extensive, so that nonextensive factors arise in the energy. Here we introduce a constraint from fluctuation theory [10] that allows simulations of the Ising model to scan between these scenarios. Of course the instantaneous energy remains extensive: the constraint simply slows down transitions from low-energy states, so that the resulting time-averaged energy per particle depends on block size. In contrast, according to Boltzmann's definition entropy is not an average, instead coming from the sum over all accessible states, so that in thermal equilibrium the average entropy must always equal the instantaneous entropy. We present evidence that the optimal value of this extra constraint maximizes the fluctuations in energy, yielding a homogeneous heat capacity per particle, and an entropy that is extensive and additive. Although there is a size-dependent reduction in the average energy of small blocks, causing a heterogeneous distribution of energies, the net free energy of the entire system is reduced, indicating better agreement with thermal equilibrium. Beyond theoretical interest, such size-dependent average energies are consistent with measurements of non-resonant spectral hole burning and nonlinear dielectric response in many materials [11-13]. Indeed, these measurements are quantitatively described with no adjustable parameters by models having a heterogeneous distribution of energies, but a homogeneous heat capacity per particle [14,15].

Computer simulations utilize dynamical constraints on model systems to mimic thermodynamic behavior [16,17]. The primary constraint for Monte-Carlo simulations is detailed balance, where the ratio of transition rates between two states must equal the Boltzmann probability for the energy difference between the states; but detailed balance alone yields fluctuations per particle that depend on the size of the subsystem block [1], yielding a nonextensive heat capacity. Perhaps the resulting nonextensive entropy is unavoidable, coming from a non-Boltzmann formula for the entropy. Alternatively, Hill's small-system thermodynamics may provide a less extreme approach for treating the finite-size thermal effects [18]. Although originally developed for isolated nanoparticles and individual molecules, this "nanothermodynamics" has been adapted to explain the independently relaxing regions in bulk materials [19,20]. Indeed, measurements showing heterogeneous dynamics demonstrate that nanoscale regions are statistically independent during relaxation, so that their entropies must be additive. The additional constraint that we investigate, which comes from standard fluctuation theory, limits the correlation between distinct degrees of freedom. Specifically, the constraint separates steps having no net alignment of nearest neighbors (constant interaction energy) from steps having no net alignment of the entire block (constant average interaction energy), thus slowing the dynamics of low-energy states, and lowering the average energy. We show that for the Ising model, this constraint yields homogeneous fluctuations, so that the specific heat and resulting entropy per particle do not depend on the size of the block.

We start with the Ising model for a system of binary degrees of freedom ("particles") on a simple cubic lattice of $N$ sites. The particle at site $i$ may be "up" ($\sigma_i$=+1) or "down" ($\sigma_i$=−1), which simulates two allowed states of a magnetic spin. (Other systems that can be related to the Ising model include structural glasses, where $\sigma_i$=+1 may correspond to a locally-favored bond-ordered structure, while $\sigma_i$=−1 may correspond to a long-ranged density-ordered structure [21].) The local field at site $i$ is $H_i=\Sigma_{<ij>}\sigma_j$, where the sum is over all six nearest neighbors. The Ising Hamiltonian is $\mathcal{H} = -\frac{1}{2} J \sum_{i=1}^{N} \sigma_i H_i$ .

Here $J$>0 for ferromagnetic interactions, the sum is over all $N$ sites in the system, and the factor of ½ removes double counting. Now consider a block of $n$ sites within the larger system, with energy $U= -\frac{1}{2} J\sum_{i=1}^{n} \sigma_i H_i$ and alignment $M=\sum_{i=1}^{n} \sigma_i$ . (We reemphasize that $H_i$ always includes all six nearest neighbors, even at the edges of each block, so that a block is defined only by the number of particles over which $U$ and $M$ are averaged.) As in Landau's theory for phase transitions, the average interaction energy as a function of alignment $\langle U(M)\rangle$ is approximately parabolic, with its maximum at $M$=0 where the slope is zero. Suppose a Monte-Carlo step inverts the particle at site $i$. If $H_i$=0 the instantaneous change in interaction energy is zero; whereas if $M$=0 the average change in interaction energy is zero [because $\langle \Delta U(0)\rangle=0$]. We find that when these two processes are separated by a constraint, the Ising model exhibits a minimum energy and homogeneous entropy, consistent with the assumptions of equilibrium nanothermodynamics and the analysis of nonlinear response measurements.

Justification for constraining the changes in interaction energy comes from fluctuation theory that requires distinct fluctuations about thermal equilibrium to be uncorrelated. The constraint is found from a Taylor-series expansion of the entropy for a small part of a large system. For magnetic variables equation 112.5 from Landau and Lifshitz [10] gives $\langle \Delta T \Delta M\rangle=0$, so that fluctuations in $T$ and $M$ must be statistically independent. Local temperature fluctuations come from changes in the interaction energy. For example, using the Creutz model for Ising spins in contact with a local heat bath [22], the average temperature is $\langle T\rangle=4J/[k_B \ln(1+4J/\langle K/n\rangle )]$, where $\langle K/n\rangle$ is the average kinetic energy per particle and $k_B$ is Boltzmann's constant. Conservation of energy yields $\Delta K=-\Delta U$, so that $\Delta T=[k_B T^2/(\langle K/n\rangle^2+4J\langle K/n\rangle)](-\Delta U/n)$, and the constraint becomes $\langle \Delta U \Delta M\rangle=0$.

Although most MC simulations have $<\Delta U \Delta M>=0$ when averaged over long times, we find better agreement with the postulate of extensive entropy using a constraint that gives $\Delta U \Delta M=0$ locally, for most steps. Because $\Delta M \neq 0$ for every successful MC step, the constraint requires that $\Delta U=0$ for most steps. As mentioned above, in Landau's theory $<U(M)>$ is approximately parabolic with its maximum at $M=0$. Thus, if $M \neq 0$ the slope of the average interaction energy is non-zero, giving a net correlation between $\Delta U$ and $\Delta M$ unless there is no change in the local interaction energy for each step, $\Delta U \equiv 0$. In other words, if $M \neq 0$ then a particle flips only if $H_i=0$. Changes in interaction energy are allowed only if the block fluctuates to $M=0$, where the slope of $<U(M)>$ is zero, giving negligible correlation between $\Delta U$ and $\Delta M$. In other words, if $M=0$ then particles in the block may flip even if $H_i \neq 0$, subject only to the constraint of detailed balance. Here we approximate the probability that a block fluctuates to $M=0$ by $p \sim \exp(\Delta S_M/k_B)$, where $\Delta S_M$ is the change in alignment entropy. Using the binomial coefficient for the multiplicity of binary degrees of freedom in a block, with Stirling's formula expanded about $M=0$, the alignment entropy can be written as $S_M = k_B [n \ln 2 - ½M^2/n]$. Thus, an approximate expression for the probability that a block fluctuates to zero alignment during a particle flip is:

$$p=\exp[- ½\, g\, M^2 / n]. \qquad \text{Eq. (1)}$$

Here $g$ is a parameter that governs the strength of the fluctuation-theory constraint. Specifically, for $g=0$ there is no additional constraint, so that the dynamics is governed entirely by detailed balance. (Here we use the standard Metropolis algorithm.) However, if $g>>1$, a particle flips only if the local field or block alignment is exactly zero, $H_i=0$ or $M=0$. We now describe the behavior of the Ising model as a function of $g$.

Figure 1 shows (a) the average interaction energy per particle ($\langle U/J\rangle/n$) and (b) the specific heat [$\langle(\Delta U/k_B T)^2\rangle/n$] as a function of the block size ($n$) over which they are averaged. The simulation is performed by subdividing a 30x30x30 lattice ($N = 27{,}000$) with periodic boundaries into 140 blocks ranging in size from $n$=6 to 1080 particles, then averaging over $10^6$ sweeps. Because every site in every block has the same number of nearest neighbors with the same value for $J$, the heterogeneous behavior of each block comes solely from intrinsic correlations. The two sets of data are from simulations at two temperatures, $T/T_C$=1.66 and 1.22, where $T_C$=4.51$J/k_B$ is the critical temperature of the simple-cubic Ising model. The solid lines have $g$=0, showing the usual result that $\langle U/J\rangle/n$ is independent of $n$, and $\langle(\Delta U/k_B T)^2\rangle/n$ increases with increasing $n$. As $g$ increases, the energy per particle decreases while the fluctuations in energy increase, until $g$>1 where the trend reverses. Near $g$=1, $\langle U/J\rangle/n$ is minimized and $\langle(\Delta U/k_B T)^2\rangle/n$ is maximized. More importantly, near $g$=1, $\langle(\Delta U/k_B T)^2\rangle/n$ becomes essentially independent of block size (dashed lines), so that the entropy per particle is homogeneous and additive. Although the average interaction energy depends on block size for $g$>0, size-dependent *average* energies are expected in nanothermodynamics.

Figures 2 (a) and (b) show the average interaction energy per particle and specific heat as a function of constraint parameter. For all different block sizes (solid symbols), $\langle U/J\rangle/n$ has a minimum and $\langle(\Delta U/k_B T)^2\rangle/n$ has a maximum near $g$=1. The open symbols show a similar $g$ dependence for the net behavior of the entire system, $N$=27,000. The approximate symmetry of the properties about $g$~1 indicates that the constraint does not artificially bias the simulations. Although the exact value at which the extrema occur depends on temperature and block size, the range of optimal $g$ is quite narrow near $g$=1, consistent with the binomial coefficient that yields Eq. (1). Thus, Eq. (1) with $g$=1 is a reasonable approximation for the probability that a fluctuation allows a change in interaction energy at these temperatures. Another approximation used in this investigation is that the distribution of blocks is fixed. The true thermal equilibrium comes from blocks that fluctuate in size, forming a dynamic distribution that eventually yields a homogeneous average interaction energy (albeit at a lower energy than if there was no heterogeneity), and an increase in entropy. The thermal equilibrium distribution is found theoretically from the generalized ensemble in nanothermodynamics [19,20]. The fixed distribution used here is sufficient to show that Eq. (1) with $g$~1 can be used to minimize $<U/n>/J$ and make $<(\Delta U/k_B T)^2>/n$ essentially independent of $n$.

Figures 3 (a) and (b) show the average interaction energy per particle and specific heat as a function of normalized alignment. To reduce statistical errors, this simulation was performed by subdividing the 30x30x30 lattice into 216 blocks, each having $n$=125 particles. The solid lines from $g$=0 show the usual parabolic curvature of Landau's theory, centered about $M$=0. The dashed lines from $g$=1 show that this constraint: (a) lowers the average energy and steepens its slope, and (b) increases the fluctuations in energy and flattens its curvature. The flattened curvature indicates improved agreement with the postulate of nanothermodynamics that similar blocks inside a larger system should have the same entropy per particle, even if they have different alignments. The inset shows the temperature dependence of the free energy of the entire lattice, $F = <U> - TS$, normalized by the temperature and total number of particles. [Entropy is found by integrating from its known value at infinite temperature, $S/k_B = N\ln(2) - \int_0^{1/T} <(\Delta U/k_B T)^2> T\,d(1/T)$ .] The additional constraint given by Eq. (1) with $g$=1 lowers the net free energy of the entire lattice for all temperatures $T_C<T<\infty$. For $T<T_C$, the optimal constraint is expected to change when the symmetry about $M$=0 is broken.

Size-dependent interaction energies, as shown in Figs. 1(a) and 2(a), provide an explanation for the heterogeneity measured on nanometer length scales in the dynamics of many materials [23-25]. Measurements and analysis of non-resonant spectral hole burning in electric [11], magnetic [12], and mechanical responses [13] demonstrate that excess energy added to selected degrees of freedom remains localized for times that may exceed thousands of seconds. A mean-field cluster model [19,20] and Landau theory [21,26] for heterogeneity, based on nanothermodynamics, predicts a $1/n$ dependence for the energy reduction from fluctuations, as shown by the dashed curves in Fig. 1 (a). The model and theory provide an explanation for the non-exponential relaxation, non-Arrhenius activation, and non-classical critical scaling found in many materials. The constraint investigated here, which conserves interaction energy during local fluctuations, gives a fundamental mechanism for the distribution of average energies used in the mean-field cluster model and theory, and for the homogeneous specific heat used to analyze non-resonant spectral hole burning.

In summary, although the entropy per particle in most Monte-Carlo simulations of interacting particles depends on the block size used for thermal averaging, we have shown that for the Ising model this size dependence can be removed by adding a constraint. The constraint, which comes from separating second-order changes in entropy during fluctuations, is added to the linear term contained in detailed balance. The additional constraint increases the thermal fluctuations of small systems, allowing the entropy to be extensive, homogeneous, and additive, consistent with a fundamental assumption of nanothermodynamics. The constraint lowers the net free energy by subdividing a large system into an ensemble of independently fluctuating blocks, providing an explanation for the dynamical heterogeneity found in many materials.

We thank Ernst Bauer, Ranko Richert, and John Shumway for helpful comments. The Fulton High Performance Computing Initiative provided technical support. Additional support for this research was provided by the Army Research Office.

References

[1] K. Binder, *Z. Phys. B Cond. Matt.* **43**, 119 (1981).

[2] R. H. Swendsen, *J. Stat. Phys.* **107**, 1143 (2002).

[3] T. Kodama, H. T. Elze, C. E. Aguiar and T. K. Koide, *Europhys. Lett.* **70**, 439 (2005).

[4] G. Wilk and Z. Wlodarczyk, *Physica A* **376**, 279 (2007).

[5] C. Tsallis, *J. Stat. Phys.* **52**, 479 (1988).

[6] C. Anteneodo and C. Tsallis, *Phys. Rev. Lett.* **80**, 5313 (1998).

[7] A. Cho, *Science* **297**, 1268 (2002).

[8] T. L. Hill, *Thermodynamics of Small Systems* (Dover, New York, 1994).

[9] T. L. Hill, *Nano Lett.* **1**, 111 (2001).

[10] L. D. Landau and E. M Lifshitz, *Statistical Physics*, Course of Theoretical Physics, vol. 5 3rd ed. chapter XII (Pergamon Press, Oxford, 1980).

[11] B. Schiener, R. Böhmer, A. Loidl, and R. V. Chamberlin, *Science* **274**, 752 (1996).

[12] R. V. Chamberlin, *Phys. Rev. Lett.* **83**, 5134 (1999).

[13] X. F. Shi and G. B. McKenna, *Phys. Rev. Lett.* **94**, 157801 (2005).

[14] R. V. Chamberlin, B. Schiener and R. Böhmer, *Mat. Res. Soc. Symp. Proc.* **455**, 117 (1997).

[15] R. Richert and S. Weinstein, *Phys. Rev. Lett.* **97**, 095703 (2006).

[16] M. E. J. Newman and G. T. Barkema, *Monte Carlo Methods in Statistical Physics*, (Oxford Press, 1999).

[17] R. V. Chamberlin and K. J. Stangel, *Phys. Lett. A* **350**, 400 (2006).

[18] R. V. Chamberlin, *Science* **298**, 1172 (2002).

[19] R. V. Chamberlin, *Phys. Rev. Lett.* **82**, 2520 (1999).

[20] R. V. Chamberlin, *Nature* **408**, 337 (2000).

[21] M. R. H. Javaheri and R. V. Chamberlin, *J. Chem. Phys.* **125**, 154503 (2007).

[22] M. Creutz, *Phys. Rev. Lett.* **50**, 1411 (1983).

[23] O. Steinsvoll and T. Riste, *Hyperfine Interactions* **31**, 267 (1986).

[24] M. D. Ediger, *Ann. Rev. Phys. Chem.* **51**, 99 (2000).

[25] R. Richert, *J. Phys. Cond. Mat.* **14**, R703 (2002).

[26] R. V. Chamberlin, *Phys. Lett. A* **315**, 313 (2003).

Figure Captions

Figure 1. (a) Average interaction energy per particle and (b) specific heat as a function of block size at two temperatures. Simulations are performed on a system of $N$=27,000 particles with periodic boundary conditions. The system is subdivided into 140 blocks, ranging in size from $n$=6 to 1080. The solid curves come from the standard Metropolis algorithm. The symbols come from simulations using the values of the constraint parameter ($g$) given in the legend. The dashed curves show that for $g$=1, $\langle U/n \rangle \sim 1/n$ and $\langle (\Delta U)^2/n \rangle \sim$constant. Error bars are shown for the $g$=1 data.

Figure 2. (a) Average interaction energy per particle and (b) specific heat as a function of constraint parameter at $T/T_C$=1.44. The solid symbols are from simulations for various blocks containing the number of particles given in the legend. Error bars are shown for the $n$=12 and $n$=96 data. The lines are a guide to the eye. The open squares show the net behavior of the entire system, $N$=27,000.

Figure 3. (a) Average interaction energy per particle and (b) specific heat as a function of alignment per particle at three temperatures. Simulations are performed on a system of $N$=27,000 particles subdivided into 216 blocks of $n$=125 particles. The solid lines are from the standard Metropolis algorithm ($g$=0), while the dashed lines have $g$=1. The inset shows the temperature dependence of the free energy per particle for the entire system, for two different block sizes, comparing g=1 to g=0. The error bars (shown for g=0) are smaller than the width of the line except at the lowest temperature.

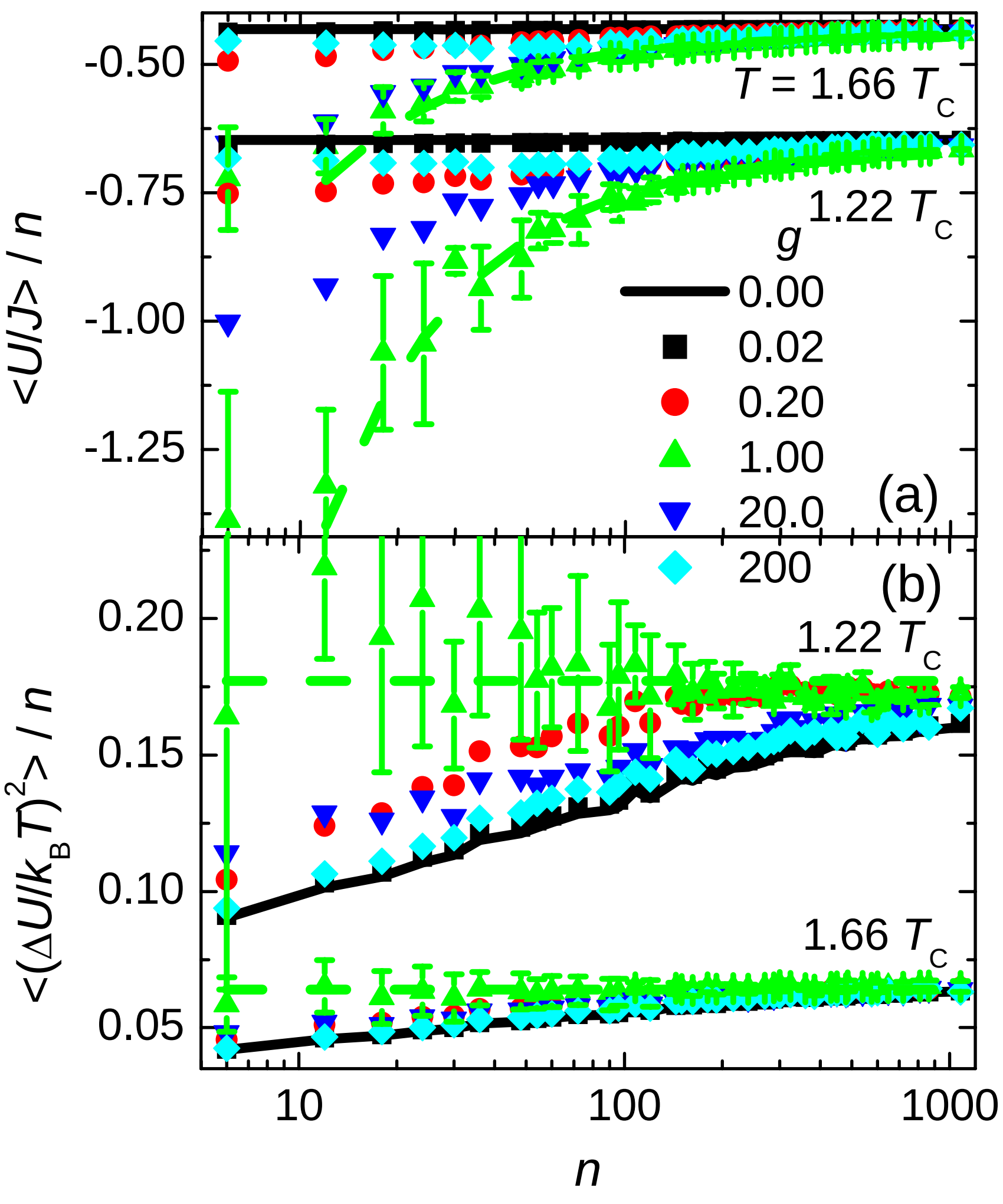

Fig. 1

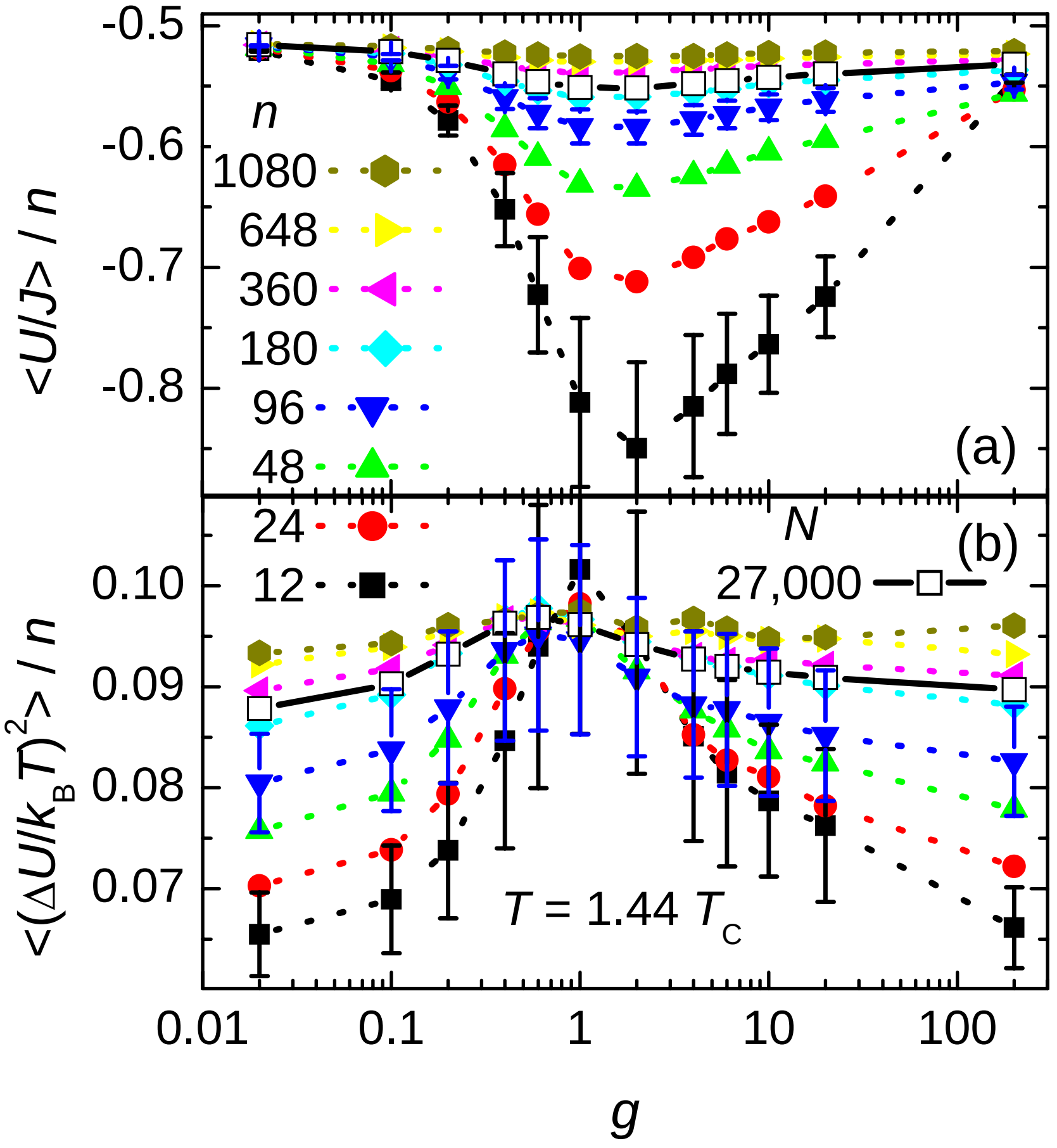

Fig. 2

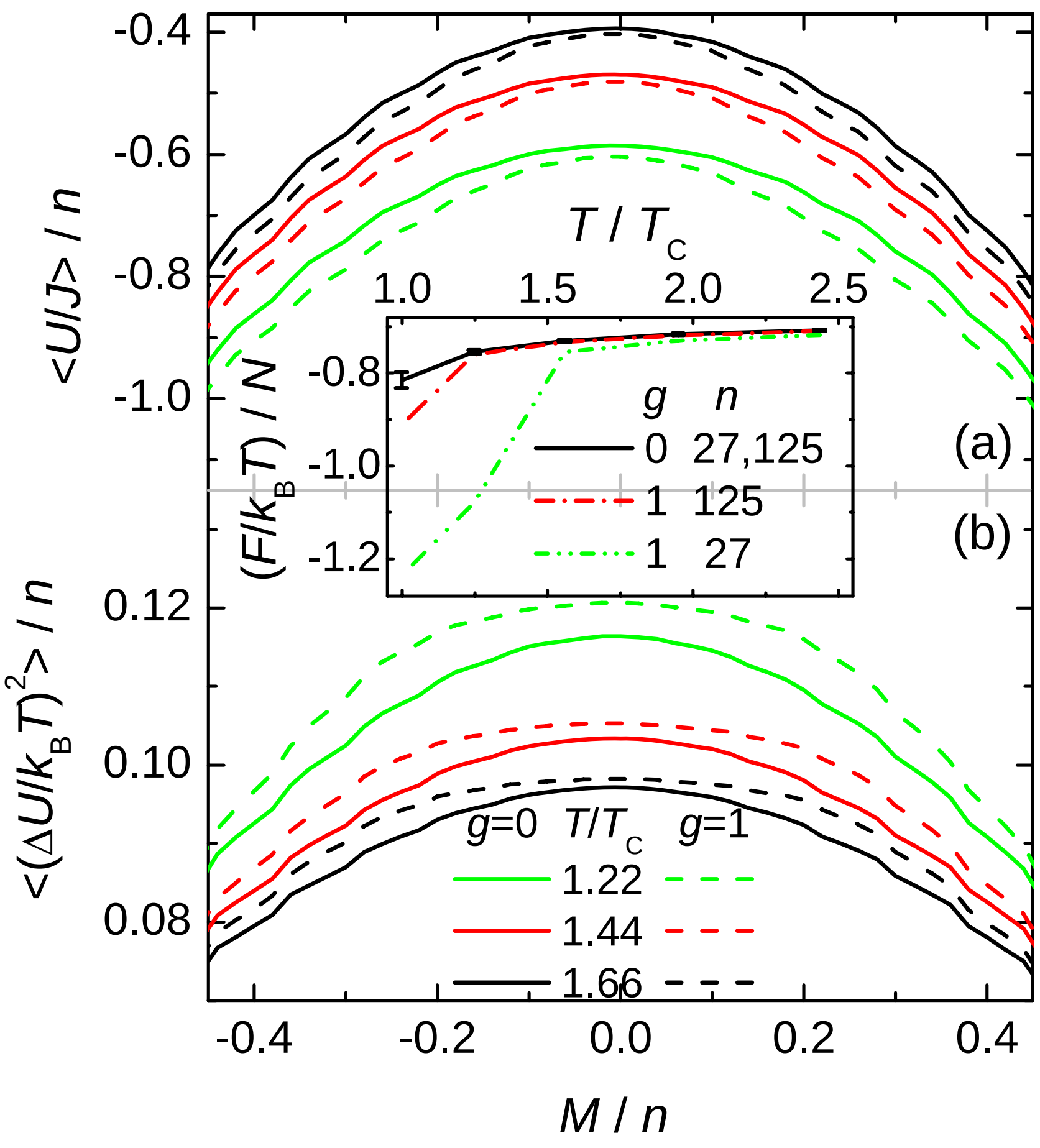

Fig. 3